\documentclass[amstex,useAMS,epsf,amssymb,usenatbib, twocolumn]{mnras}
\usepackage{epsfig}
\usepackage{graphicx}
\usepackage{color}
\usepackage{amssymb,amsmath}
\usepackage{multicol}
\usepackage{soul}
\usepackage[normalem]{ulem}
\usepackage{url}
\usepackage{hyperref}
\usepackage{breakurl}
\usepackage{tabularx}

\newcolumntype{Y}{>{\centering\arraybackslash}X}

\newcommand{\fermilat}{{\it Fermi}-LAT}
\newcommand{\gray}{$\gamma$-ray}
\newcommand{\grays}{$\gamma$ rays}

\def\araa{ARA\&A}             
\def\apj{ApJ}                 
\def\apjl{ApJ}                
\def\apjs{ApJS}               
\def\aap{A\&A}                
\def\mnras{MNRAS}             
\def\prd{Phys.~Rev.~D.}
\def\physrep{Phys.~Rep.}   
\def\ssr{Space Science Reviews}

\title[CR hadrons in AGN-inflated bubbles]{Confinement and diffusion time-scales of CR hadrons in AGN-inflated bubbles}

\author[D. A. Prokhorov \& E. M. Churazov]{D. A. Prokhorov$^{1}$\thanks{E-mail:phdmitry@gmail.com}
and E. M. Churazov$^{2, 3}$
\\
~\\
$^{1}$ School of Physics, Wits University, Private Bag 3, WITS-2050,
Johannesburg, South Africa
\\
$^{2}$ Max Planck Institute for Astrophysics,
Karl-Schwarzschild-Strasse 1, 85741 Garching, Germany
\\
$^{3}$ Space Research Institute (IKI), Profsouznaya 84/32, Moscow
117997, Russia
}

\date{Accepted .....
      Received ..... ;
      in original form .....}

\pagerange{\pageref{firstpage}--\pageref{lastpage}} \pubyear{2017}

\begin{document}

\maketitle

\begin{abstract}
While rich clusters are powerful sources of X-rays, \gray{} emission
from these large cosmic structures has not been detected yet. X-ray radiative energy losses in the central regions of relaxed galaxy clusters are so strong that one needs to consider special sources of energy, likely AGN feedback, to suppress catastrophic cooling of the gas. We consider a model of AGN feedback that postulates that the AGN supplies the energy to the gas by inflating bubbles of relativistic plasma, whose energy content is dominated by cosmic-ray (CR) hadrons. If most of these hadrons can quickly escape the bubbles, then collisions of CRs with  thermal protons in the intracluster
medium (ICM) should lead to strong \gray{} emission, unless fast diffusion of CRs removes them from the cluster.
Therefore, the lack of detections with modern \gray{} telescopes sets limits on the confinement time of CR hadrons in bubbles and CR diffusive propagation in the ICM.
\end{abstract}


\section{Introduction}

X-ray radiation observed from clusters of galaxies is emitted by
hot, low-density, and metal-enriched plasmas filling the space
between galaxies \citep[for a review, see][]{Sarazin1986}.
 We focus attention on relaxed galaxy clusters which are not
disturbed by ongoing mergers with other groups (or clusters) of
galaxies. The measured X-ray surface brightness and derived plasma
density profiles of such clusters are sharply
centrally peaked and decrease with radius, while the derived plasma
temperature profiles have minima at the centres
\citep[for a review, see, e.g.,][and references therein]{Peterson2006}.
These central low-temperature dense regions are coined as cool cores
\citep[][]{Molendi2001} (formerly known as ``cooling flows'' ) and galaxy clusters having cool cores are
called cool-core clusters. If no external energy sources are
present in cool cores, the radiative losses in clusters' centres would lead to gas cooling well below X-ray temperatures in a fraction of  the Hubble time. The absence of
the ``low temperature'' emission lines in the spectra of
cool-core clusters \citep[e.g.,][]{Tamura2001} and the lack of intense star formation suggest that external energy
sources for plasma heating do exist in the cool cores. A likely explanation is that the energy produced by the
central active galactic nuclei (AGN) balances radiative
losses of the X-ray emitting plasma \citep[e.g.,][]{Churazov2000, McNamara2000}.
The activity of the central AGN manifests itself via the presence of X-ray ``cavities'' (or bubbles) inflated by relativistic outflows from the AGN that are present in majority of cool core clusters \citep[for a review, see][]{Fabian2012, Bykov2015}. These bubbles are typically bright in the radio band \citep[e.g.,][]{Bohringer1993}, implying that relativistic electrons (and, perhaps, positrons) are present inside them. The bubbles rise buoyantly in cluster atmospheres and pure mechanical interaction can provide efficient energy transfer from the bubble enthalpy to the gas \citep[e.g.,][]{Churazov2000,Churazov2001,Churazov2002}. These buoyancy arguments suggest that the energy supplied by the AGN matches approximately the gas cooling losses. This conclusion has been confirmed by the analysis of many systems that differ in X-ray luminosity by several orders of magnitude \citep[e.g.,][]{Birzan2004,Hlavacek2012}.

So far there is no firm conclusion on the content of the observed bubbles.
We only know that there are relativistic electrons
that are responsible for the synchrotron radiation of the bubbles. Very hot, but still non-relativistic, matter could also make significant contribution to the bubbles' pressure and the most direct way to probe this component observationally is to use the Sunyaev-Zeldovich effect \citep[][]{Pfrommer2005,Prokhorov2010}. It is also possible, that relativistic protons dominate the energy content of the bubbles \citep[][]{Dunn2004}, similar to the situation in the Galaxy, where CR protons dominate electrons by a factor of the order of 100.
The signature of the presence of CR hadrons in AGN-inflated bubbles
might be established through \gray{} emission produced due to a
decay of neutral pions created in the inelastic interactions of CR hadrons with thermal protons.
Inside the bubbles this mechanism is unlikely to be important, since the density of protons is small, and the relativistic hadrons have first to escape from bubbles into the ICM \citep[][]{Mathews2009}.

Diffuse \gray{} emission from galaxy clusters has not been
detected yet even with the modern \gray{} pair conversion telescope,
\fermilat{}
\citep[e.g.,][]{Huber2013, Ackermann2014, Prokhorov2014} or
the  Cherenkov telescopes including H.E.S.S.
\citep[e.g.,][]{HESSclusters},
MAGIC
\citep[e.g.,][]{MAGICLIMITS, magic2016}, and
VERITAS
\citep[][]{VERITASCOMA}. However, the X-ray brightest cool-core
clusters, Perseus and Virgo, in which AGN-inflated bubbles are very prominent,
do contain bright central \gray{} emitting sources. These
\gray{} sources detected with \fermilat{} and Cherenkov telescopes
are identified with the cores of radiogalaxies, NGC 1275
\citep[][]{NGC1275FERMI, NGC1275MAGIC} and M 87 \citep[][]{M87FERMI,
M87VERITAS, M87MAGIC}, respectively. The Perseus cluster is expected
to be one of the most promising cool-core clusters to constrain the
pressure of CR hadrons due to its high X-ray flux and high central
concentration of the expected \gray{} flux \citep[][]{MAGICLIMITS,
magic2016}. The observations of the Perseus cluster with MAGIC
result in the upper limit on the average CR-to-thermal pressure
ratio of $<2\%$. The tight bounds on the ratio of CR
proton-to-thermal pressures of $\lesssim1.5\%$ have also been
obtained with \fermilat{} through the stacking of $\simeq 50$ galaxy
clusters of smaller angular sizes \citep[e.g.,][]{Huber2013,
Ackermann2014, Prokhorov2014}.

In this paper we show that the flux upper limits provided by the modern \gray{}
telescopes can strongly constrain the models of both the CR
hadron confinement in AGN-inflated bubbles and diffusive propagation of
CR hadrons in the ICM. We will come to the conclusion that if CR
hadrons are injected by AGN jets in bubbles, these CR hadrons have to be confined
in buoyantly rising bubbles much longer than the sound crossing time
of cool cores.

The structure of the paper is the following. In \S\ref{sec:estimates} we provide the order-of-magnitude estimates and rough scaling of the \gray{} flux. In \S\ref{sec:toyModel} we present a more elaborate, but still very straightforward model for \gray{} flux computation. \S\ref{sec:results} summarizes our findings.

\section{Order-of-magnitude \gray{} flux estimates}
\label{sec:estimates}

An overall concept of our model is the following; We assume that the
energy provided by the AGN feedback matches the gas cooling losses
$L_X$. We further assume that relativistic protons dominate in the
feedback energy balance. Thus, the total production rate of CR
protons by the AGN is known and equal to the observed X-ray
luminosity coming from the region within the cooling radius. This
injection of relativistic protons by itself is not associated with
the \gray{} flux, since the protons are initially confined within
the bubbles. The bubbles rise buoyantly through the cluster
atmosphere with the velocity  $u_{\mathrm{adv}}\sim \eta
c_{\mathrm{s}}$, where $c_{\mathrm{s}}$ is the sound speed in the
ICM, and $\eta$ is in the range 0.2-0.5. During the rise, the CR
protons experience adiabatic losses, thus reducing the total energy
associated with them. A fraction of protons can escape from the
bubble as it rises and enter the ICM. The rate of escape is
characterised by the confinement time $t_{\mathrm{conf}}$, so that
the fraction of protons escaping during a short interval $\Delta t$ is
$\Delta t/t_{\mathrm{conf}}$. Once the relativistic proton is in the
ICM, it diffuses in space with a diffusion coefficient $D$.
The proton in the ICM can collide with the thermal
protons and generate \gray{} flux.

Apart from diffusive propagation of CR protons in the ICM, the CRs
can stream relative to the bulk plasma along magnetic field lines
\citep[see, e.g.][]{Wentzel1974}. The streaming of the CRs with
respect of the ICM excites magnetohydrodynamic waves, on which the
CRs scatter. Generation of waves and scattering on them limits CR
streaming speeds to the speed of the waves, which is the Alfv\'{e}n
speed, but if the waves are strongly damped, highly
super-Alfv\'{e}nic streaming is possible. If the streaming can
efficiently transport the CRs down to their pressure gradient from
the central regions of cool-core clusters, it can be an alternative
mechanism to quench diffuse $\gamma$-ray emission, while dissipation
of the waves can provide additional heating of the ICM plasma in
cool cores. We do not include CR streaming in this study. The effect
of CR streaming  in the context of galaxy clusters was studied by
e.g., \citet[][]{Pfrommer2013, Wiener2013, Ruszkowski2017}.

Thus, in the frame of the model considered here, the most important (unknown)
parameters affecting the \gray{} flux are the confinement time $t_{\mathrm{conf}}$
and the diffusion coefficient $D$. Below we begin our analysis by considering the
most extreme values of these parameters that should lead to the maximal \gray{} flux.

\subsection{Maximal \gray{} flux}
The \gray{} flux $F_\gamma$ (in units of ${\rm phot\;s^{-1}\;cm^{-2}}$) coming from a volume $V$ with the density of thermal protons ${n}_{\mathrm{pl}}$ can be estimated as
\begin{equation}
F_{\gamma}\simeq\frac{q_{\gamma} \epsilon_{\mathrm{CRp}} V
n_{\mathrm{pl}} }{4\pi d^2},
\label{Eq2}
\end{equation}
where $q_{\gamma}$ is the \gray{} emissivity normalised to the unit CR
hadron energy density is given by \citet[][]{Drury1994} and
$\epsilon_{\mathrm{CRp}}$ is the energy density of CR hadrons.
 It is obvious that the maximal \gray{} flux can
be expected if all hadrons are released immediately into the ICM
(without being trapped inside the bubble, i.e.,
$t_{\mathrm{conf}}\rightarrow 0$) and they do not diffuse outside
the central region (i.e., $D\rightarrow 0$). Of course this is not a
realistic setup, but it serves as an absolute upper limit on the
\gray{} flux. The rate of energy losses by relativistic protons weakly depends
on proton energy $E\gtrsim 100\;{\rm GeV}$. We assume
\begin{equation}
\frac{\mathrm{d}\ln E}{\mathrm{d}t}\approx 3.85\times10^{-16}\left(\frac{n_{\mathrm{pl}}}{{\mathrm{cm}}^{-3}}\right)~{\rm s^{-1}},
\label{eq:loss}
\end{equation}
see, e.g., \cite{Krakau2015}. The total energy released by the AGN over the life-time of the clusters, which we assume to be comparable to the Hubble time $t_H$, is $\epsilon_{\mathrm{CRp}} V\sim L_X t_{H}$. If the plasma density $n_{\mathrm{pl}}\geq6\times10^{-3}$ cm$^{-3}$ then Eq.~(\ref{eq:loss}) implies that the life time of relativistic protons $\displaystyle t_{\mathrm{loss}}=\left( \frac{\mathrm{d}\ln E}{\mathrm{d}t}\right)^{-1}\sim \frac{8\times10^{7}}{n_{pl}} {\rm yr} $ is shorter than $t_H$. This condition is satisfied in the centres of all cool core clusters. In this case, $\epsilon_{\mathrm{CRp}} V\sim L_{\mathrm{X}} t_{\mathrm{loss}}$  and the plasma density ${n}_{\mathrm{pl}}$
term in Eq.\ref{Eq2} is cancels out
\begin{equation}
F_{\gamma,{\rm max}}\sim\frac{q_{\gamma} L_{\mathrm{X}}}{4\pi d^2 \left (\frac{1}{n_{\mathrm{pl}}}\frac{\mathrm{d}\ln E}{\mathrm{d}t}\right)}\sim \frac{}{4\pi d^2} \frac{L_X}{78~{\rm erg}},
\label{eq:fmax}
\end{equation}
where we use $q_{\gamma}=4.9\times10^{-18}$ s$^{-1}$
erg$^{-1}$ cm$^3$ (H-atom)$^{-1}$ (for $\Gamma=-2.2$ from
\citealt{Drury1994}) for the production of gamma-rays above 1~TeV.
For the Perseus cluster, using the X-ray luminosity within
the cooling radius $L_X$ and distance to the cluster $d$ from \citet[][]{Birzan2012}, the estimated \gray{} flux above 1 TeV equals to $1\times10^{-11}$
cm$^{-2}$s$^{-1}$ and exceeds the MAGIC upper limit (for the \textit{point-like} model),
$4\times10^{-14}$ cm$^{-2}$s$^{-1}$ above 1 TeV
\citep[][]{magic2016}, by about 250 times. Therefore, in the absence of streaming losses, the observational constraints
permit us to rule out the case corresponding to the short
confinement time of CR hadrons and the slow CR diffusion in the ICM.

\subsection{Pure diffusion case}
The same arguments can be easily extended to the case of a finite
diffusion coefficient, while keeping the confinement time small
$t_{\mathrm{conf}}\rightarrow 0$. The \gray{} flux will be smaller than predicted by Eq.~(\ref{eq:fmax})
if diffusion can transport a large fraction of CR protons from the central region into lower density outer plasmas, where $t_{loss}\gtrsim t_{H}$. In this case,  $\epsilon_{\mathrm{CRp}}
V\sim L_X t_{H}$, but now the effective value of $n_{\mathrm{pl}}$
in Eq.~(\ref{Eq2}) depends on the size of the region occupied by
diffused CR protons. Consider a stationary solution of the diffusion
equation with a steady point source of cosmic rays at the centre. In
the central region the constant diffusive flux through any radius
$r$ implies that   $\epsilon_{\mathrm{CRp}}(r)\propto  r^{-1}$.
Assuming that ${n}_{\mathrm{pl}}(r)$ declines with radius slower
than $r^{-2}$ one can conclude that the integral $\displaystyle \int
\epsilon_{\mathrm{CRp}}(r) {n}_{\mathrm{pl}}(r) r^2 dr$ is dominated
by the largest radii $r_{\mathrm{max}}$. One can choose
$r_{\mathrm{max}}\sim \sqrt{D t_{\mathrm{H}}}$. Thus,
\begin{equation}
F_{\gamma,{\rm dif}}\sim\frac{q_{\gamma} L_{\mathrm{X}} t_{\mathrm{H}}
n_{\mathrm{pl}}(r_{\mathrm{max}})}{4\pi d^2}.
\label{eq:fdif}
\end{equation}
If $r_{\mathrm{max}}\sim 1$~Mpc, the MAGIC flux upper limit for
the \textit{point-like} model of \gray{} emission from the
Perseus cluster is inapplicable since it was obtained within the
reference radius of 0.15$^{\circ}$ (or 200 kpc at z=0.0179).
The MAGIC flux upper limit obtained within the virial radius for the
\textit{extended} model is a factor of 15 larger than that for
the \textit{point-like} model and, given that predicted \gray{} flux slowly increases with radius, the constraints
obtained on a basis of the flux from the cool-core region are tighter.

It takes about $t_{\mathrm{D}}\sim R^2_{\mathrm{core}}/D$ for CR
protons to spread out by diffusion from the cluster cool-core.
During this time interval, the energy released in
CR hadrons in the central region is $L_{\mathrm{X}} t_{\mathrm{D}}$.
The \gray{} flux produced owing to CR protons not having a
sufficient time to diffuse beyond the cool-core region is
\begin{equation}
F_{\gamma}\sim\frac{q_{\gamma} L_{\mathrm{X}} t_{\mathrm{D}} n_{\mathrm{pl}}(R_{\mathrm{core}})}{4\pi d^2}. \label{eq:fpure}
\end{equation}

From Eq. (\ref{eq:fpure}), one can find that to satisfy the MAGIC
flux upper limit for the \textit{point-like} model, the diffusion
coefficient has to be $D>5\times10^{31}$ cm$^2$s$^{-1}$.
\subsection{Pure advection case}
For the finite $t_{\mathrm{conf}}$ the calculations are slightly more complicated, since one has to take into account adiabatic losses during the buoyant rise of the bubbles. This is done in \S\ref{sec:toy} below. Here we consider a special (and rather artificial) case of zero diffusion ($D=0$) and very long $t_{\mathrm{conf}}$, such that only a small fraction of protons escape the bubble over its entire life time. In this approximation we can assume that the energy content of the rising bubble is changing only due to adiabatic expansion.  The momentum of a relativistic proton $p$ inside the bubbles changes as $p=p_0 \left ( P(r)/P_0\right )^{(\gamma-1)/\gamma}$, where $p_0$ is the initial particle momentum, $P_0$ and $P(r)$ are the initial and current pressure (at radius $r$) of the ICM and $\gamma=4/3$.
Accordingly, the number of relativistic particles, $N$, producing \grays{} above given energy $E_\gamma=p_0c$ is
\begin{eqnarray}
N(r)\propto N_0  \left ( \frac{P_0}{P(r)}\right )^{\frac{\gamma-1}{\gamma}(1+\Gamma)},
\end{eqnarray}
where $N_0$ is the initial number of such particles with $p>p_0$ and the CR hadron power-law
spectrum, $\propto p^{\Gamma}$. With this correction the expression for the \gray{} flux can be written as

\begin{equation}
F_{\gamma,{\rm adv}}\sim \frac{q_{\gamma} L_X}{4\pi d^2} \int \frac{N(r)}{N_0}
n_{\mathrm{pl}}(r)\frac{ t_{\mathrm{min}}}{u_{\mathrm{adv}}t_{\mathrm{conf}}}  \mathrm{d}r,
\label{eq:fadv}
\end{equation}
where $t_{\mathrm{min}}$ is the minimal time among $t_{\mathrm{H}}$ and $\displaystyle t_{\mathrm{loss}}\sim \frac{8\times10^{7}}{n_{\mathrm{pl}}(r)} {\rm yr}$.
The integral in expression (\ref{eq:fadv}) is dominated by the range
of radii where the radial density profile is shallower than $r^{-1}$,
i.e., the inner $\sim$100 kpc of the Perseus cluster, especially
taking into account the adiabatic losses that further reduce the
contribution of the outer regions. Using $r\sim 100$~kpc and
$n_{\mathrm{pl}}\sim 7\times10^{-3}~{\rm cm^{-3}}$ for estimates
one gets an order-of-magnitude constraint on $\displaystyle
\frac{r}{u_{\mathrm{adv}} t_{\mathrm{conf}}}\lesssim
5\times10^{-3}$, or, equivalently, $t_{\mathrm{conf}}\gtrsim
4\times10^{10}$~yr.

\subsection{Most constraining galaxy clusters}

In this Section, we use simple estimates to pre-select the most promising
clusters in terms of the possible constraints on the confinement time and the diffusion coefficient.

Most of the gas properties of cool-core
clusters are known from X-ray observations. Since \gray{} diffuse
emission from cool-core clusters has not been detected yet, both
the amount and distribution of CR hadrons are unknown. We apply a
scaling law to select cool-core clusters expected to be bright
in \grays{}. To deduce the \gray{} scaling law, we use four basic
assumptions:

(a) The production rate of CR hadrons is proportional to the X-ray luminosity
of the cooling radius region,
$\dot{N}_{\mathrm{CR}}\propto L_{\mathrm{X}}$ (energy constraint);

(b) The X-ray luminosity within the cooling radius is
$L_{\mathrm{X}}\propto \int^{R_{\mathrm{cool}}}_{0} n^2_{\mathrm{pl}} \mathrm{d}^3 r$,
where $R_{\mathrm{cool}}$ and $n_{\mathrm{pl}}$ are the cooling radius and
the plasma number density in the central region, respectively;

(c) The \gray{} flux is proportional to the product of CR hadron and thermal plasma
number densities,
$F_{\gamma}\propto\dot{N}_{\mathrm{CR}} n_{\mathrm{pl}}/d^2$;

(d) The life time of CR protons due to hadronic losses is inversely proportional to the plasma density.

Using these assumptions, we expect the approximate \gray{} scaling relation,

\begin{equation}
F_{\gamma}\propto\frac{L_{\mathrm{X}}}{d^2}.
\label{Eq1}
\end{equation}

Putting the values of parameters for the sample of cool-core
clusters taken from \citet[][]{Birzan2012}, we find that the highest
\gray{} flux is expected from the central region of the Perseus
cluster. The second highest \gray{} flux is expected from the
central region of the Ophiuchus cluster (which is about 4 times
lower than that from the Perseus cluster), and the third highest
\gray{} flux is expected from M87 (which is
about 5 times lower than the expected \gray{} flux from
the Perseus cluster). Other clusters expected to be bright in
\grays{} are Abell 2029, Abell 2199, Abell 478, Centaurus,
2A 0335+096, and Abell 1795 (those fluxes are about 10 times lower
than that expected from the Perseus cluster).

Observational upper limits on the \gray{} fluxes are different for
nearby galaxy clusters because these limits depend on the presence
of \gray{} emitting central radio galaxies in cool cores, on how
sensitive various \gray{} telescopes are, and on how strong the
Galactic foreground emission is in the direction of galaxy clusters.
To investigate which cool-core cluster can provide us with the
tightest constraints on the CR hadron confinement time and on the CR
diffusion coefficient, one needs to calculate the flux upper limits
scaled to the flux above the same energy (above 1 TeV in this
paper). These scaled fluxes have to be computed assuming the
production of $\gamma$ rays via a neutral pion decay.
To scale the flux from 1 GeV to 1 TeV, we used Eq. 19 from
\citet[]{pfrommer2004} derived in the framework of the Dermer's
model \citep[][]{Dermer1986} and assumed that the values of the CR hadron power-law
spectral index, $\Gamma$, are $-2.2$ and $-2.5$. The upper limit on
the \gray{} flux (Table \ref{T2}) from the Perseus cluster above 1
TeV was derived for the \textit{point-like} model by
\citet[][]{magic2016}. The flux upper limits above 1 GeV for
cool-core clusters (including Abell 478, Abell 2199, and Abell 1795)
derived from the \textit{Fermi}-LAT data are taken from
\citet[][]{Ackermann2014}, while the flux upper limits above 1 GeV
for the Ophiuchus cluster, 2A 0335+096, Centaurus, and Abell 2029
are taken from \citet[][]{Selig}. The \gray{} flux from the Virgo
cluster is that of M87 at a low flux state \citep[][]{M87MAGIC}. The
scaled flux upper limits above 1 TeV for these cool-core clusters
are also shown in Table \ref{T2}.

\begin{table}
 \centering
\caption{The scaled flux upper limits (in cm$^{-2}$s$^{-1}$) above 1 TeV for the selected
cool-core clusters (and the estimator values)}
\begin{tabular}{| c | c | c | c | c |}
\hline Cluster name & Scaled flux limit, & Scaled flux limit, \\ &
$\Gamma=-2.2$ & $\Gamma=-2.5$ \\ \hline
Perseus & $3.8\times10^{-14}$ (1.00) & $3.8\times10^{-14}$ (0.34)\\
Abell 1795 & $1.2\times10^{-14}$ (0.24) & $1.0\times10^{-15}$ (1.00)\\
Abell 478 & $2.0\times10^{-14}$ (0.18)  & $1.8\times10^{-15}$ (0.68)\\
Ophiuchus & $6.2\times10^{-14}$ (0.15) & $5.4\times10^{-15}$ (0.57)\\
Abell 2199 & $3.9\times10^{-14}$ (0.11) & $3.4\times10^{-15}$ (0.42)\\
Centaurus & $3.2\times10^{-14}$ (0.11) & $2.8\times10^{-15}$ (0.42)\\
Abell 2029 & $4.7\times10^{-14}$ (0.08)  & $4.1\times10^{-15}$ (0.32)\\
2A 0335+096 & $4.1\times10^{-14}$ (0.07)  & $3.5\times10^{-15}$ (0.31)\\
Virgo (M87) & $4.6\times10^{-13}$ (0.01)  &  --\\
\hline
\end{tabular}
\label{T2}
\end{table}

To select the most constraining
targets for studying the CR hadron confinement in the bubbles,
we introduce an estimator defined as
the ratio of the expected flux\footnote{computed using the scaling
relation from Eq. \ref{Eq1}} to the upper limit of the
scaled observational \gray{} flux. The values of the estimator for the selected
cool-core clusters are shown in parentheses in Table \ref{T2}. We
normalised the estimator to its highest value for each of the values of
the CR hadron spectral index,  $\Gamma$. We found that the Perseus
cluster corresponds to the highest value of  the estimator, if
$\Gamma=-2.2$. For the softer power-law index, $\Gamma=-2.5$, the
situation is different and the cool-core clusters with the scaled
fluxes obtained from the \textit{Fermi}-LAT data give estimator
values similar to (or even higher than) that derived for the Perseus cluster
by using the MAGIC observations.

\section{Modelling of \gray{} emission produced by escaped CR hadrons}
\label{sec:toy}

In this Section, we introduce the toy model
that combines constraints on the CR hadron escape from AGN-inflated bubbles
and CR diffusive propagation in the ICM. In the framework of this model, we
will compute the \gray{} flux expectation and will set limits on the
confinement and diffusion time scales of CR hadrons.

Both the confinement and the diffusion control transport of CR
hadrons in cool-core clusters. Confined CR hadrons can be
transported inside the bubbles from
the cluster centre to outer cluster regions. The longer the confinement
time, the fewer CR hadrons escape from the bubbles into the ICM when
passing the densest cooling region. After CR hadrons escape their parent
bubbles, they then diffuse and fill the intracluster medium. In this
case CR hadron propagation is specified by the diffusion
coefficient. After time, $\tau$, the CR hadrons diffuse a distance
$l\sim (D \tau)^{1/2}$. If the diffusion coefficient is sufficiently
large, e.g. $D\sim10^{31}$ cm$^2$ s$^{-1}$, CR hadrons injected into
the ICM within the cooling region cannot be retained in the cooling
region for $10^{10}$ years. Therefore, both the long confinement
time and the fast diffusion cause a decrease in the containment of
CR hadrons within the cooling radius. Apart from confinement and
diffusion, hadronic losses also affect the containment of CR hadrons
within the cooling radius, because CR hadrons cannot remain in the
cooling radius longer than their life time due to hadronic losses.
Beyond the cooling radius, the life time of CR protons due to
hadronic losses is much longer because of the lower plasma density
in the outer regions. Although CR protons can exist in the outer
regions for the Hubble time, \gray{} emission from the outer
regions is comparably small because of the same reason.

\subsection{Toy model}
\label{sec:toyModel}

To calculate the \gray{} fluxes from the cool-core galaxy clusters
expected due to the interactions of CR hadrons that escaped from rising
bubbles with the ICM, we introduce a toy model having the
properties:

({i}) Mechanical work done by jets to inflate bubbles and that done
by rising bubbles during their adiabatic expansion in the ICM
compensates X-ray radiative energy losses within the cooling radius;

({ii}) Isolated galaxy clusters are assumed to be spherically
symmetric, i.e., plasma temperature and density profiles only depend
on a radius. These profiles are determined from X-ray observations
and to simplify the case we ignore the evolution of these profiles
in time;

({iii}) We also assume that AGN-inflated bubbles are filled with CR
hadrons and that CR pressure inside the bubbles
equal ambient plasma pressure. Thus, the CR hadron energy density
in the bubbles is determined solely by the radial plasma pressure profile;

({iv}) AGN-inflated bubbles rise with a constant sub-sonic speed,
$u_{\mathrm{adv}}$, along the radius-vector from the centre of a
cool-core cluster. This determines the spatial
profile of the CR hadron source function which drops as the square
of the radius from the centre of a cool-core cluster;

({v}) The CR hadron escape from the bubbles (in other words,
the injection of CR hadrons in the ICM) is described by the source term in the CR
diffusion equation. For the sake of simplicity, we assume that the
time of CR hadron confinement $t_{\mathrm{conf}}$, does not depend
on energy;

({vi}) The number density and low energy bound of CR hadrons in
each bubble change with time and, therefore, depend on the radial
position of the bubble. To calculate how these parameters change
with time, we solved the system of equations
\begin{eqnarray}
\frac{N_{\mathrm{CR}} \epsilon}{3V}&=&P_{\mathrm{pl}} \\
\frac{\partial N_{\mathrm{CR}}}{\partial t}&=& \frac{-N_{\mathrm{CR}}}{t_{\mathrm{conf}}} \\
\epsilon&=&\epsilon_{0} \left(\frac{V}{V_{0}}\right)^{1-\gamma},
\end{eqnarray}
where $N_{\mathrm{CR}}$, $\epsilon$, $V$, and $P_{\mathrm{pl}}$ are
the number of cosmic rays, the energy per particle in a bubble, and
the volume of a bubble, and the plasma pressure as a function of
radius from the centre of the cluster, respectively;

({vii}) We assume that the CR hadron spectrum has a power-law
shape. The change of a low energy bound of CR hadrons
with time leads to the injection of CR hadrons into the ICM at
different low energy bounds. This would affect the shape of the CR
hadron spectrum at low energies. We do not include this effect
and consider the model under the condition that
the initial minimal kinetic energy of CR hadrons, $E_{\mathrm{kin,
0}}$, is lower than a typical energy of CR hadrons,
$E_{\mathrm{p}}\gtrsim10 E_{\gamma\mathrm{, min}}$, producing
\grays{} above the minimal observed energy $E_{\gamma\mathrm{,
min}}$ (e.g., if $E_{\gamma\mathrm{, min}}=1$ GeV then
$E_{\mathrm{kin, 0}}\lesssim10$ GeV). Note that the minimal kinetic
energy of CR hadrons inside the bubbles decreases with time due to
the bubble adiabatic expansion. Therefore, if the latter condition
is satisfied at the initial moment it will also be satisfied as AGN
bubbles evolve in time. To compute the energy budget the minimal
kinetic energy of a few GeV was assumed, thus the \gray{}
emissivity values used in this paper are conservative.

{({viii}) Under the assumption that escaped CR hadrons are well mixed
with the ICM, CR hadrons undergo proton-proton hadronic interactions
with an ICM plasma. Owing to these interactions CR hadrons produce
neutral pions, each decaying into two \grays{}. The expected \gray{}
flux is expressed as $F_{\gamma}= \sigma_{\mathrm{pp}} c
f(E_{\gamma}) \int n_{\mathrm{cr}}(r) n_{\mathrm{pl}}(r) r^2
\mathrm{d}r/d^2$, where $\sigma_{\mathrm{pp}}$ is the inelastic
p-p cross section,
and $f(E_{\gamma})$ is the function of $\gamma$-ray energy
which is obtained from a simple fitting form given by \citet[Sect.
3.2.2,][]{pfrommer2004}. Note that the CR hadron number density,
$n_{\mathrm{cr}}(r)$, is derived by solving the diffusion equation
and the plasma number density profile, $n_{\mathrm{pl}}(r)$, is
taken from X-ray observations.}

({ix}) The transport of CR hadrons through the ICM is diffusive. The
CR diffusion coefficient, $D$, is also assumed not to depend on
energy. This corresponds to the case of passive advective transport
in a turbulent flow with the diffusion coefficient of
$D=v_{\mathrm{turb}} \lambda_{\mathrm{turb}}/3\sim10^{29}$ cm$^2
s^{-1}$, where $v_{\mathrm{turb}}\sim100$ km s$^{-1}$ and
$\lambda_{\mathrm{turb}}\sim10$ kpc are the turbulent velocity and
coherence length, respectively \citep[see, e.g.,][]{pfrommer2004}.
In addition to a transport in cluster turbulent flows, wave-CR
interactions result in CR propagation with the energy dependent CR
diffusion coefficient. In the latter case, the CRs are scattered by
the magnetic fluctuations of the microscopic scales. The random
component of magnetic field with a Kolmogorov spectrum of
inhomogeneities leads to the diffusion coefficient of $D\propto
E^{1/3}$ \citep[see, e.g.,][in the context of galaxy
clusters]{Volk1996}. If the microscopic CR transport is dominant,
then the constraints obtained in the framework of the toy model can
be re-estimated\footnote{Note that the higher the CR diffusion
coefficient is, the more extended the distribution of CR protons
becomes. In the case of the diffusion coefficient increasing with
energy, the lowest
 $\gamma$-ray flux is expected for the highest value of the
diffusion coefficient (due to a plasma density profile decreasing
with radius). If the integral $\gamma$-ray flux at energies higher
than $E_{\mathrm{min}}$ is determined by CR protons with energies
between $10 E_\mathrm{min}$ and $10^4 E_\mathrm{min}$ \citep[see,
e.g.,][for $\gamma$-ray spectra produced at p-p
collisions]{Kelner2006}, then the constraints obtained on the diffusion 
coefficient, $D$, in the framework of the toy model can conservatively be converted to the
constraints on the CR diffusion coefficient at energy of $10
E_\mathrm{min}$ by dividing them by a factor of $\sim \left(
10^4/10\right)^{1/3}\sim10$ in the case of a Kolmogorov spectrum of
magnetic field fluctuations.}. The life time of CR protons due to
hadronic losses is inversely proportional to the plasma density and
only weakly depends on energy in the high energy regime
\citep[][]{Krakau2015}. To simplify the model, we assume that the
the CR hadron life time due to hadronic losses does not depend on
energy (in this case the energy term cancels off in the diffusion
equation). We solve the diffusion equation numerically using the
Crank-Nicolson method.

These basic properties of the toy model allow us to calculate the
gamma-ray fluxes expected from cool-core galaxy clusters if CR
hadron escape from the bubbles into the ICM takes place.
This toy model has two free parameters which are the confinement
time of CR hadrons in the bubble, and the CR diffusion coefficient
in the ICM.

\subsection{Constraints on CR hadron confinement and diffusion}
\label{sec:results}

\begin{figure*}
\centering
  \begin{tabular}{@{}cc@{}}
    \includegraphics[angle=90, width=.45\textwidth]{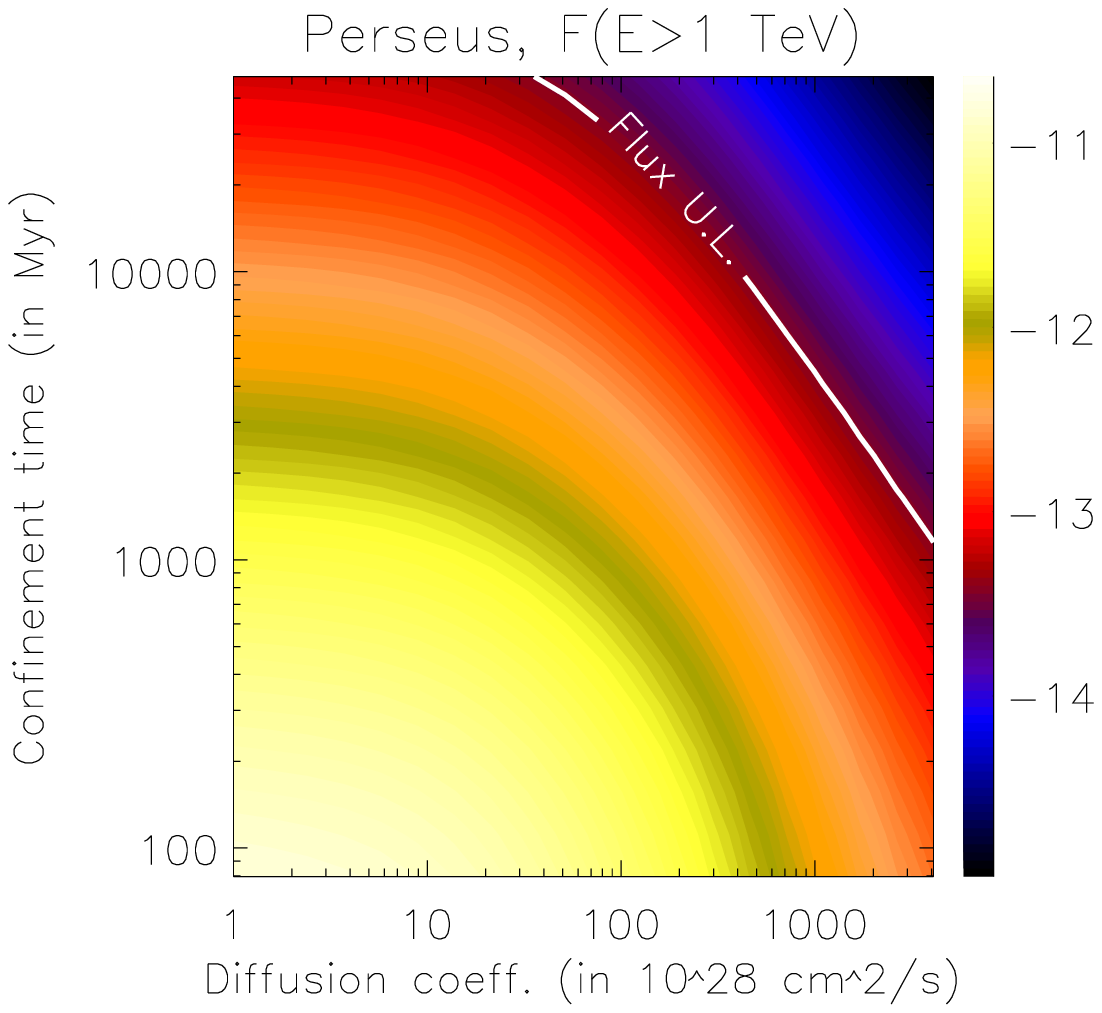}
    & \ \ \ \ \ \ \ \ \
    \includegraphics[angle=90, width=.45\textwidth]{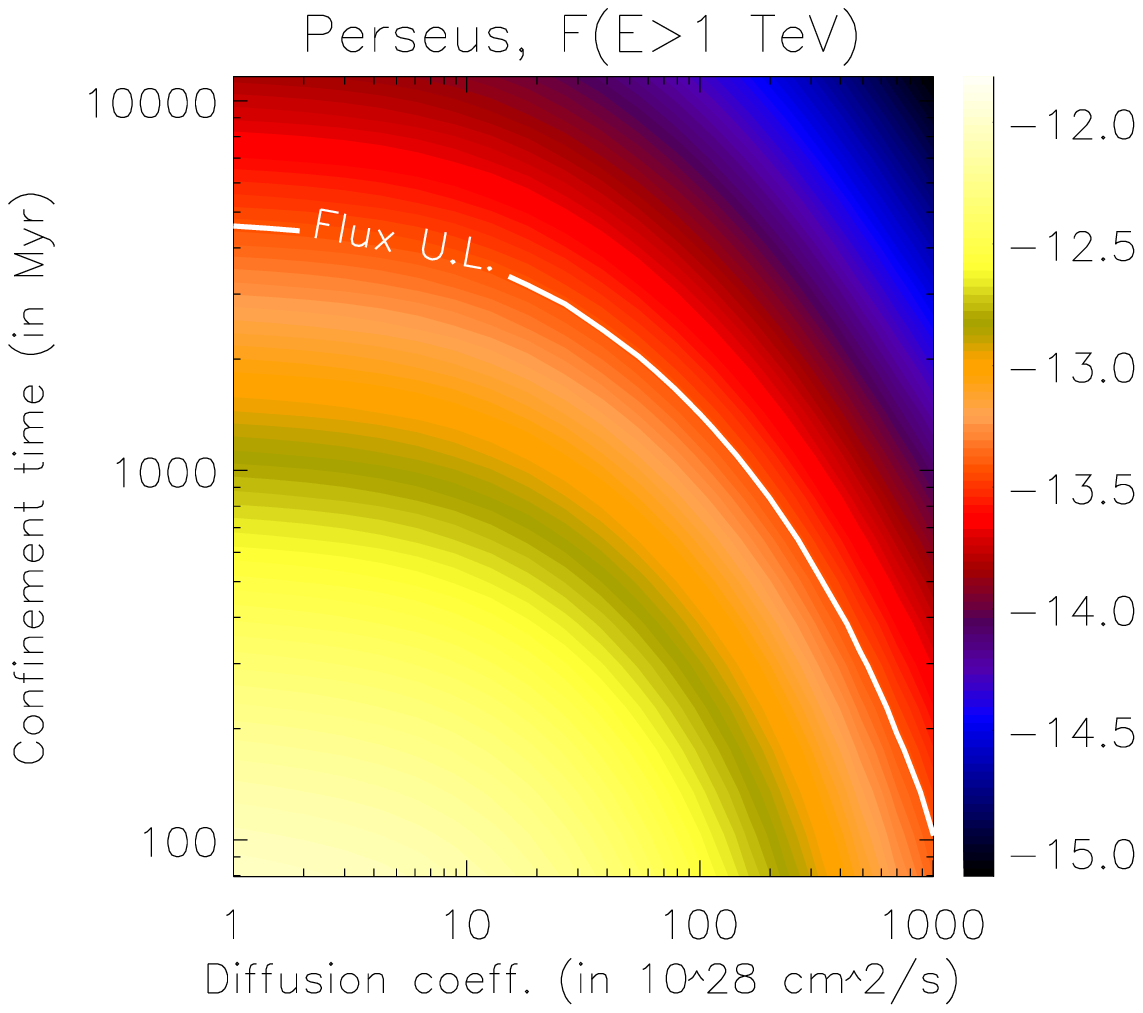} \\
   \end{tabular}
  \caption{Logarithm of the $\gamma$-ray flux in ph cm$^{-2}$ s$^{-1}$ and 
constraints on the parameter space ($t_{\mathrm{conf}}$, $D$)
obtained for the Perseus cluster for the spectral index values of
$\Gamma=-2.2$ (left panel) and $\Gamma=-2.5$ (right panel)}
\label{F1}
\end{figure*}

\begin{figure*}
\centering
  \begin{tabular}{@{}cc@{}}
    \includegraphics[angle=90, width=.45\textwidth]{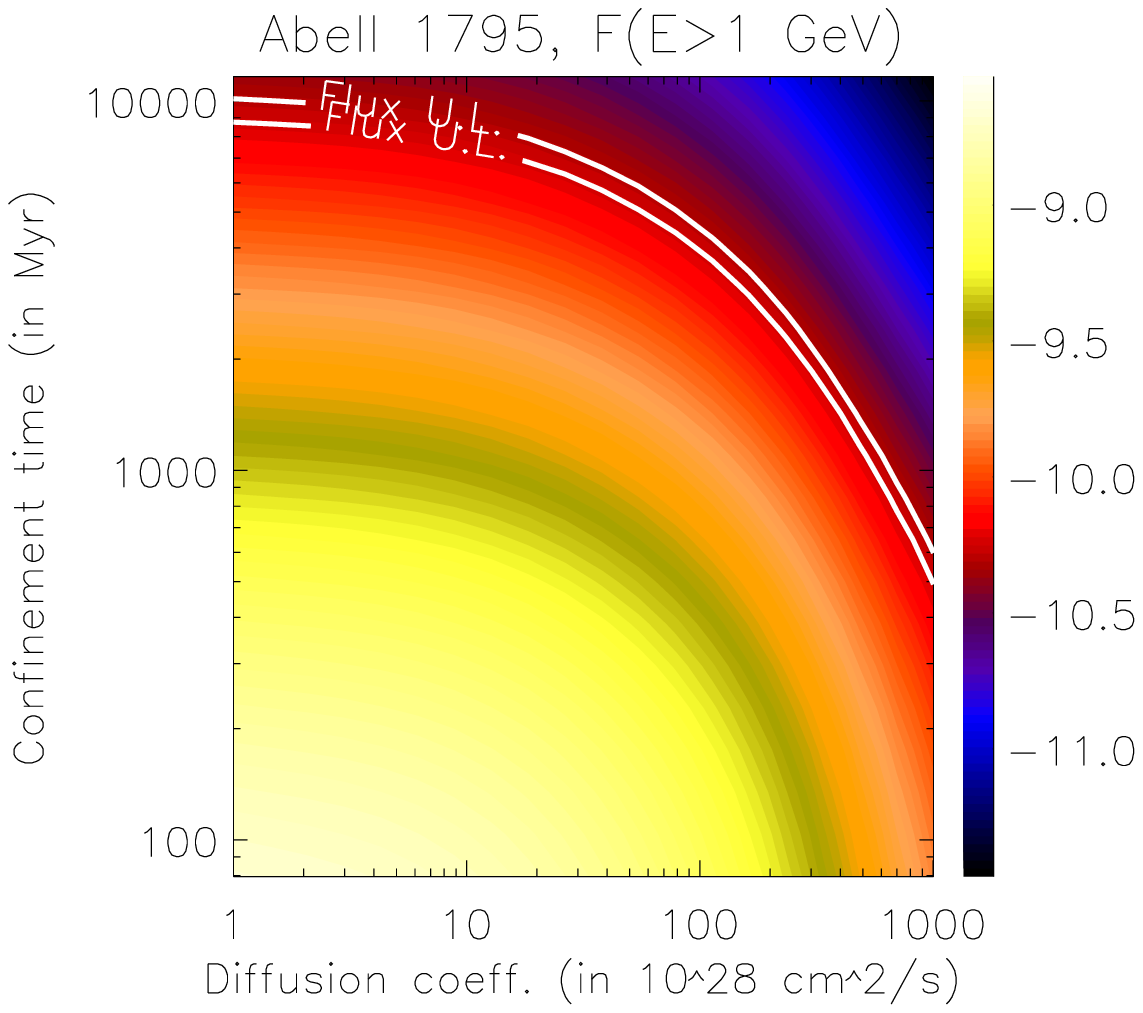}
    & \ \ \ \ \ \ \ \ \ \
    \includegraphics[angle=90, width=.45\textwidth]{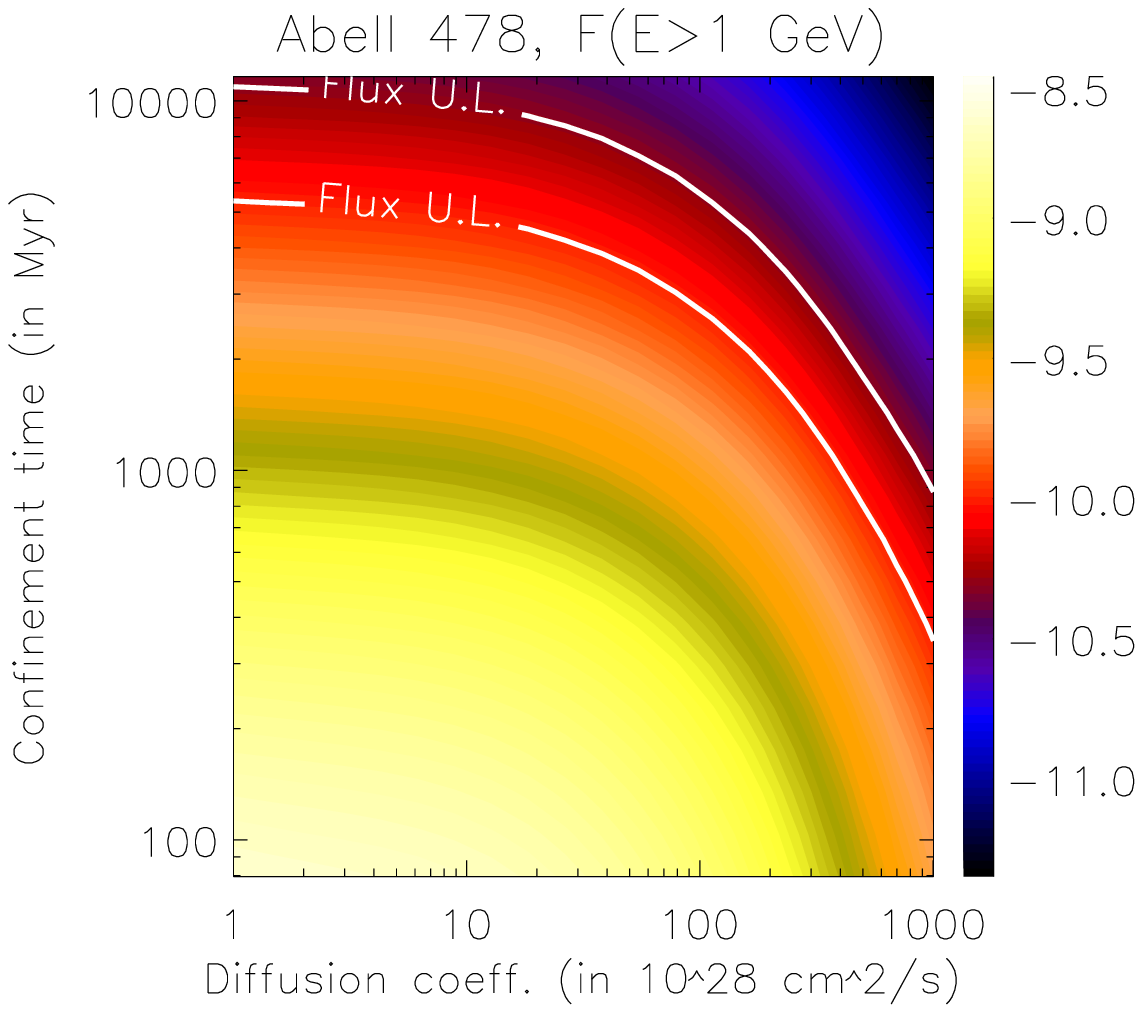} \\
   \end{tabular}
  \caption{Logarithm of the $\gamma$-ray flux in ph cm$^{-2}$ s$^{-1}$ and constraints 
on the parameter space ($t_{\mathrm{conf}}$, $D$)
obtained for Abell 1795 (left panel) and Abell 478 (right panel) for
the spectral index value of $\Gamma=-2.5$.} \label{F2}
\end{figure*}

In this Section, we present the results of numerical computations
obtained in the framework of the toy model. We are pursuing the goal
of computing the expected \gray{} fluxes from cool-core
clusters as functions of the two parameters,
the CR confinement time, $t_\mathrm{conf}$, and the CR diffusion
coefficient, $D$. We select the three cool-core clusters,
Perseus, Abell 1795, and Abell 478, expected to be the most
constraining (see Table \ref{T2}) for the investigation.

Observational \gray{} flux upper limits are used
to constrain the two free parameters of the model. The gas parameters
of the model, including the radial profiles of plasma density
and pressure in the ICM are taken from \cite{Churazov2003,
Zhuravleva2013} for the Perseus cluster and from \cite{Vikhlinin2006}
for the Abell 1795 and Abell 478 clusters.
For the Perseus cluster, the \gray{}
flux is estimated within the cylindrical region of a reference radius,
$0.15^{^{\circ}}$, to make comparisons with the flux upper limit
obtained for the \textit{point-like} model with MAGIC (see Sect. 2.2).
The other two cool-core clusters under investigation,
Abell 1795 and Abell 478 located is about 270 and 385 Mpcs from Earth
are point-like sources for \textit{Fermi}-LAT.
For Abell 1795 and Abell 478, the \gray{} fluxes are estimated within the
region of virial radii to make comparisons with the flux upper
limits obtained with \textit{Fermi}-LAT.

We found that the MAGIC observations of the Perseus cluster provide
the tightest constraints if the spectral index is hard,
$\Gamma=-2.2$. Figure \ref{F1} shows the constraints obtained for
the Perseus cluster for the spectral index values of
$\Gamma=-2.2$ and $\Gamma=-2.5$. Both these constraints are based on
the observations above 1 TeV. The constraints are much tighter for
$\Gamma=-2.2$ than those for $\Gamma=-2.5$. It is because of a much
higher production rate (normalised to the CR hadron energy density)
of \grays{} above 1 TeV for the former hadron spectral index (see
Table 1 from \citet[][]{Drury1994}). This shows that AGN-inflated
bubbles keep CR hadrons for a time, $t_{\mathrm{conf}}>10^{10}$ yr,
if $D<3\times10^{30}$ cm$^2$ s$^{-1}$ and $\Gamma=-2.2$.

We checked the compatibility of the obtained constraints with
those estimated in Sect. \ref{sec:estimates} for the Perseus
cluster. The computed maximal flux, $1.5\times10^{-11}$
ph~cm$^{-2}$~s$^{-1}$, is close to that estimated above. The
computed constraints obtained for the pure diffusion and pure
advection cases are also similar to those estimated above.
Therefore, the order-of-magnitude \gray{} flux estimates
are compatible with the \gray{} fluxes obtained in the
framework of the toy model.

The results of the computations in the case of $\Gamma=-2.5$ for
Abell 1795 and Abell 478 are shown in Fig. \ref{F2}. We found that
the constraints obtained using the \textit{Fermi}-LAT flux upper
limits for these cool-core clusters (bottom white lines in the
panels of Fig. \ref{F2}) are similar to or even tighter than those
obtained for the Perseus cluster (if $\Gamma=-2.5$). It agrees with
the result obtained by means of the estimators in Sect.
\ref{sec:estimates} (see Table 1). The constraints obtained for
Abell 1795 and Abell 478 for the spectral index of $\Gamma=-2.2$ are
similar to those are obtained for these clusters if $\Gamma=-2.5$.
To strengthen the constraints, we applied the \gray{} flux upper limit
obtained through the stacking of 50 cool-core clusters
\citep[e.g.,][]{Huber2013, Ackermann2014, Prokhorov2014}. The upper
white curves in the panels of Fig. \ref{F2} show this \gray{} flux
upper limit. We found that the bubbles keep CR hadrons for a time,
$t_{\mathrm{conf}}>5\times10^{9}$ yr, if $D<1\times10^{30}$
cm$^2$s$^{-1}$ and $\Gamma=-2.5$. Note that the \gray{} flux upper
limit for Abell 1795 is close to that obtained through the stacking
(the scaled flux upper limit of Abell 1795 in Table \ref{T2} is the
tightest).

The sound crossing times of the cooling regions in the Perseus
cluster, Abell 1795, and Abell 478 are about 100 Myr. Therefore,
CR hadrons are confined in the bubble over times much longer
than the sound crossing time under the condition that the diffusion
coefficient, $D$, is less than $10^{31}$ cm$^2$ s$^{-1}$ (see Figs.
\ref{F1}, \ref{F2}).

We cross-checked and confirmed these constraints using an
alternative method of computations based on the solution of the
nonhomogeneous diffusion equation by means of the Green function.
The Green function was obtained by applying the Duhamel's principle
to the textbook solution of the homogeneous diffusion equation in
spherical coordinates. The advantage of the main method, the
Crank-Nicolson finite difference method, used in this paper is that
it allows the inclusion of hadronic losses in a simple way, while hadronic
losses were not included using the alternative method.

\section{Conclusions}

AGN are the likely heat source of the cool cores in relaxed clusters of galaxies. Observational evidence of the AGN interaction with the gas at the centres of cool-core clusters is the presence of bubbles in the hot surrounding medium. Mechanical power of the bubbles is sufficient to balance radiative X-ray cooling in the cool cores. If the bubbles consist of CR hadrons, \grays{} produced in collisions between CR hadrons, escaped from the bubbles, and the ICM protons can be measured by modern \gray{} telescopes. High-energy observations of the cool-core clusters
provide important insights into the physical processes of CR hadron confinement in the inflated bubbles and into CR transport models.

This paper reports results of modelling of \gray{} emission expected owing
to the escape of CR hadrons from the bubbles in cool-core clusters. The order-of-magnitude \gray{} flux estimates were performed and showed that the maximal \gray{} flux obtained for the Perseus cluster under the conditions (all hadrons are released immediately into the ICM and do not diffuse outside the central region) strongly exceeds the observed flux upper limit. The pure diffusion and pure advection cases were investigated and the order-of-magnitude limits on the diffusion coefficient and on the CR confinement time scale were derived. The \gray{} scaling relation was derived and used to
select galaxy clusters whose existing \gray{} observations are expected to
result in the tightest constraints on the parameter space of CR confinement and transport models.
The toy model was introduced to carry out more
quantitative calculations (see Sect. \ref{sec:toyModel}). This model was
applied to the selected cool-core clusters and numerical computations
using the Crank-Nicolson method were performed. Comparing the computed
\gray{} fluxes with the observed upper flux limits, the two free parameters
of the model, CR confinement time scale and diffusion coefficient,
were constrained.
CR hadrons are confined in the bubbles for a time of, $t_{\mathrm{conf}}>10^{10}$ yr, if $D<3\times10^{30}$ cm$^2$ s$^{-1}$ and $\Gamma=-2.2$,
and for a time of $t_{\mathrm{conf}}>5\times10^{9}$ yr, if $D<1\times10^{30}$ cm$^2$s$^{-1}$ and $\Gamma=-2.5$. These time scales are much longer than
the sound crossing times of the cooling regions in the investigated clusters.
These constraints are tight, but could possibly be relaxed if other physical
processes, such as streaming, complicate the simple model used in this paper.

\section{Acknowledgements}
EMC acknowledges partial support by grant No. 14-22-00271 from the
Russian Scientific Foundation.
DAP acknowledges support from the DST/NRF SKA post-graduate bursary initiative.

\bibliography{refs_v7}

\begin{thebibliography}{}
\makeatletter
\relax
\def\mn@urlcharsother{\let\do\@makeother \do\$\do\&\do\#\do\^\do\_\do\%\do\~}
\def\mn@doi{\begingroup\mn@urlcharsother \@ifnextchar [ {\mn@doi@}
  {\mn@doi@[]}}
\def\mn@doi@[#1]#2{\def\@tempa{#1}\ifx\@tempa\@empty \href
  {http://dx.doi.org/#2} {doi:#2}\else \href {http://dx.doi.org/#2} {#1}\fi
  \endgroup}
\def\mn@eprint#1#2{\mn@eprint@#1:#2::\@nil}
\def\mn@eprint@arXiv#1{\href {http://arxiv.org/abs/#1} {{\tt arXiv:#1}}}
\def\mn@eprint@dblp#1{\href {http://dblp.uni-trier.de/rec/bibtex/#1.xml}
  {dblp:#1}}
\def\mn@eprint@#1:#2:#3:#4\@nil{\def\@tempa {#1}\def\@tempb {#2}\def\@tempc
  {#3}\ifx \@tempc \@empty \let \@tempc \@tempb \let \@tempb \@tempa \fi \ifx
  \@tempb \@empty \def\@tempb {arXiv}\fi \@ifundefined
  {mn@eprint@\@tempb}{\@tempb:\@tempc}{\expandafter \expandafter \csname
  mn@eprint@\@tempb\endcsname \expandafter{\@tempc}}}

\bibitem[\protect\citeauthoryear{{Abdo} et~al.,}{{Abdo}
  et~al.}{2009a}]{NGC1275FERMI}
{Abdo} A.~A.,  et~al., 2009a, \mn@doi [\apj] {10.1088/0004-637X/699/1/31},
  \href {http://adsabs.harvard.edu/abs/2009ApJ...699...31A} {699, 31}

\bibitem[\protect\citeauthoryear{{Abdo} et~al.,}{{Abdo}
  et~al.}{2009b}]{M87FERMI}
{Abdo} A.~A.,  et~al., 2009b, \mn@doi [\apj] {10.1088/0004-637X/707/1/55},
  \href {http://adsabs.harvard.edu/abs/2009ApJ...707...55A} {707, 55}

\bibitem[\protect\citeauthoryear{{Acciari} et~al.,}{{Acciari}
  et~al.}{2008}]{M87VERITAS}
{Acciari} V.~A.,  et~al., 2008, \mn@doi [\apj] {10.1086/587458}, \href
  {http://adsabs.harvard.edu/abs/2008ApJ...679..397A} {679, 397}

\bibitem[\protect\citeauthoryear{{Ackermann} et~al.,}{{Ackermann}
  et~al.}{2014}]{Ackermann2014}
{Ackermann} M.,  et~al., 2014, \mn@doi [\apj] {10.1088/0004-637X/787/1/18},
  \href {http://adsabs.harvard.edu/abs/2014ApJ...787...18A} {787, 18}

\bibitem[\protect\citeauthoryear{{Aharonian} et~al.,}{{Aharonian}
  et~al.}{2009}]{HESSclusters}
{Aharonian} F.,  et~al., 2009, \mn@doi [\aap] {10.1051/0004-6361:200811372},
  \href {http://adsabs.harvard.edu/abs/2009A%26A...495...27A} {495, 27}

\bibitem[\protect\citeauthoryear{{Ahnen} et~al.,}{{Ahnen}
  et~al.}{2016}]{magic2016}
{Ahnen} M.~L.,  et~al., 2016, \mn@doi [\aap] {10.1051/0004-6361/201527846},
  \href {http://adsabs.harvard.edu/abs/2016A%26A...589A..33A} {589, A33}

\bibitem[\protect\citeauthoryear{{Aleksi{\'c}} et~al.,}{{Aleksi{\'c}}
  et~al.}{2012a}]{NGC1275MAGIC}
{Aleksi{\'c}} J.,  et~al., 2012a, \mn@doi [\aap] {10.1051/0004-6361/201118668},
  \href {http://adsabs.harvard.edu/abs/2012A%26A...539L...2A} {539, L2}

\bibitem[\protect\citeauthoryear{{Aleksi{\'c}} et~al.,}{{Aleksi{\'c}}
  et~al.}{2012b}]{MAGICLIMITS}
{Aleksi{\'c}} J.,  et~al., 2012b, \mn@doi [\aap] {10.1051/0004-6361/201118502},
  \href {http://adsabs.harvard.edu/abs/2012A%26A...541A..99A} {541, A99}

\bibitem[\protect\citeauthoryear{{Aleksi{\'c}} et~al.,}{{Aleksi{\'c}}
  et~al.}{2012c}]{M87MAGIC}
{Aleksi{\'c}} J.,  et~al., 2012c, \mn@doi [\aap] {10.1051/0004-6361/201117827},
  \href {http://adsabs.harvard.edu/abs/2012A%26A...544A..96A} {544, A96}

\bibitem[\protect\citeauthoryear{{Arlen} et~al.,}{{Arlen}
  et~al.}{2012}]{VERITASCOMA}
{Arlen} T.,  et~al., 2012, \mn@doi [\apj] {10.1088/0004-637X/757/2/123}, \href
  {http://adsabs.harvard.edu/abs/2012ApJ...757..123A} {757, 123}

\bibitem[\protect\citeauthoryear{{B{\^i}rzan}, {Rafferty}, {McNamara}, {Wise}
  \& {Nulsen}}{{B{\^i}rzan} et~al.}{2004}]{Birzan2004}
{B{\^i}rzan} L.,  {Rafferty} D.~A.,  {McNamara} B.~R.,  {Wise} M.~W.,
  {Nulsen} P.~E.~J.,  2004, \mn@doi [\apj] {10.1086/383519}, \href
  {http://adsabs.harvard.edu/abs/2004ApJ...607..800B} {607, 800}

\bibitem[\protect\citeauthoryear{{B{\^i}rzan}, {Rafferty}, {Nulsen},
  {McNamara}, {R{\"o}ttgering}, {Wise}  \& {Mittal}}{{B{\^i}rzan}
  et~al.}{2012}]{Birzan2012}
{B{\^i}rzan} L.,  {Rafferty} D.~A.,  {Nulsen} P.~E.~J.,  {McNamara} B.~R.,
  {R{\"o}ttgering} H.~J.~A.,  {Wise} M.~W.,   {Mittal} R.,  2012, \mn@doi
  [\mnras] {10.1111/j.1365-2966.2012.22083.x}, \href
  {http://adsabs.harvard.edu/abs/2012MNRAS.427.3468B} {427, 3468}

\bibitem[\protect\citeauthoryear{{Boehringer}, {Voges}, {Fabian}, {Edge}  \&
  {Neumann}}{{Boehringer} et~al.}{1993}]{Bohringer1993}
{Boehringer} H.,  {Voges} W.,  {Fabian} A.~C.,  {Edge} A.~C.,   {Neumann}
  D.~M.,  1993, \mn@doi [\mnras] {10.1093/mnras/264.1.L25}, \href
  {http://adsabs.harvard.edu/abs/1993MNRAS.264L..25B} {264, L25}

\bibitem[\protect\citeauthoryear{{Bykov}, {Churazov}, {Ferrari}, {Forman},
  {Kaastra}, {Klein}, {Markevitch}  \& {de Plaa}}{{Bykov}
  et~al.}{2015}]{Bykov2015}
{Bykov} A.~M.,  {Churazov} E.~M.,  {Ferrari} C.,  {Forman} W.~R.,  {Kaastra}
  J.~S.,  {Klein} U.,  {Markevitch} M.,   {de Plaa} J.,  2015, \mn@doi [\ssr]
  {10.1007/s11214-014-0129-4}, \href
  {http://adsabs.harvard.edu/abs/2015SSRv..188..141B} {188, 141}

\bibitem[\protect\citeauthoryear{{Churazov}, {Forman}, {Jones}  \&
  {B{\"o}hringer}}{{Churazov} et~al.}{2000}]{Churazov2000}
{Churazov} E.,  {Forman} W.,  {Jones} C.,   {B{\"o}hringer} H.,  2000, \aap,
  \href {http://adsabs.harvard.edu/abs/2000A%26A...356..788C} {356, 788}

\bibitem[\protect\citeauthoryear{{Churazov}, {Br{\"u}ggen}, {Kaiser},
  {B{\"o}hringer}  \& {Forman}}{{Churazov} et~al.}{2001}]{Churazov2001}
{Churazov} E.,  {Br{\"u}ggen} M.,  {Kaiser} C.~R.,  {B{\"o}hringer} H.,
  {Forman} W.,  2001, \mn@doi [\apj] {10.1086/321357}, \href
  {http://adsabs.harvard.edu/abs/2001ApJ...554..261C} {554, 261}

\bibitem[\protect\citeauthoryear{{Churazov}, {Sunyaev}, {Forman}  \&
  {B{\"o}hringer}}{{Churazov} et~al.}{2002}]{Churazov2002}
{Churazov} E.,  {Sunyaev} R.,  {Forman} W.,   {B{\"o}hringer} H.,  2002,
  \mn@doi [\mnras] {10.1046/j.1365-8711.2002.05332.x}, \href
  {http://adsabs.harvard.edu/abs/2002MNRAS.332..729C} {332, 729}

\bibitem[\protect\citeauthoryear{{Churazov}, {Forman}, {Jones}  \&
  {B{\"o}hringer}}{{Churazov} et~al.}{2003}]{Churazov2003}
{Churazov} E.,  {Forman} W.,  {Jones} C.,   {B{\"o}hringer} H.,  2003, \mn@doi
  [\apj] {10.1086/374923}, \href
  {http://adsabs.harvard.edu/abs/2003ApJ...590..225C} {590, 225}

\bibitem[\protect\citeauthoryear{{Dermer}}{{Dermer}}{1986}]{Dermer1986}
{Dermer} C.~D.,  1986, \aap, \href
  {http://adsabs.harvard.edu/abs/1986A%26A...157..223D} {157, 223}

\bibitem[\protect\citeauthoryear{{Drury}, {Aharonian}  \& {Voelk}}{{Drury}
  et~al.}{1994}]{Drury1994}
{Drury} L.~O.,  {Aharonian} F.~A.,   {Voelk} H.~J.,  1994, \aap, \href
  {http://adsabs.harvard.edu/abs/1994A%26A...287..959D} {287, 959}

\bibitem[\protect\citeauthoryear{{Dunn} \& {Fabian}}{{Dunn} \&
  {Fabian}}{2004}]{Dunn2004}
{Dunn} R.~J.~H.,  {Fabian} A.~C.,  2004, \mn@doi [\mnras]
  {10.1111/j.1365-2966.2004.08365.x}, \href
  {http://adsabs.harvard.edu/abs/2004MNRAS.355..862D} {355, 862}

\bibitem[\protect\citeauthoryear{{Fabian}}{{Fabian}}{2012}]{Fabian2012}
{Fabian} A.~C.,  2012, \mn@doi [\araa] {10.1146/annurev-astro-081811-125521},
  \href {http://adsabs.harvard.edu/abs/2012ARA%26A..50..455F} {50, 455}

\bibitem[\protect\citeauthoryear{{Hlavacek-Larrondo}, {Fabian}, {Edge},
  {Ebeling}, {Sanders}, {Hogan}  \& {Taylor}}{{Hlavacek-Larrondo}
  et~al.}{2012}]{Hlavacek2012}
{Hlavacek-Larrondo} J.,  {Fabian} A.~C.,  {Edge} A.~C.,  {Ebeling} H.,
  {Sanders} J.~S.,  {Hogan} M.~T.,   {Taylor} G.~B.,  2012, \mn@doi [\mnras]
  {10.1111/j.1365-2966.2011.20405.x}, \href
  {http://adsabs.harvard.edu/abs/2012MNRAS.421.1360H} {421, 1360}

\bibitem[\protect\citeauthoryear{{Huber}, {Tchernin}, {Eckert}, {Farnier},
  {Manalaysay}, {Straumann}  \& {Walter}}{{Huber} et~al.}{2013}]{Huber2013}
{Huber} B.,  {Tchernin} C.,  {Eckert} D.,  {Farnier} C.,  {Manalaysay} A.,
  {Straumann} U.,   {Walter} R.,  2013, \mn@doi [\aap]
  {10.1051/0004-6361/201321947}, \href
  {http://adsabs.harvard.edu/abs/2013A%26A...560A..64H} {560, A64}

\bibitem[\protect\citeauthoryear{{Kelner}, {Aharonian}  \& {Bugayov}}{{Kelner}
  et~al.}{2006}]{Kelner2006}
{Kelner} S.~R.,  {Aharonian} F.~A.,   {Bugayov} V.~V.,  2006, \mn@doi [\prd]
  {10.1103/PhysRevD.74.034018}, \href
  {http://adsabs.harvard.edu/abs/2006PhRvD..74c4018K} {74, 034018}

\bibitem[\protect\citeauthoryear{{Krakau} \& {Schlickeiser}}{{Krakau} \&
  {Schlickeiser}}{2015}]{Krakau2015}
{Krakau} S.,  {Schlickeiser} R.,  2015, \mn@doi [\apj]
  {10.1088/0004-637X/802/2/114}, \href
  {http://adsabs.harvard.edu/abs/2015ApJ...802..114K} {802, 114}

\bibitem[\protect\citeauthoryear{{Mathews}}{{Mathews}}{2009}]{Mathews2009}
{Mathews} W.~G.,  2009, \mn@doi [\apjl] {10.1088/0004-637X/695/1/L49}, \href
  {http://adsabs.harvard.edu/abs/2009ApJ...695L..49M} {695, L49}

\bibitem[\protect\citeauthoryear{{McNamara} et~al.,}{{McNamara}
  et~al.}{2000}]{McNamara2000}
{McNamara} B.~R.,  et~al., 2000, \mn@doi [\apjl] {10.1086/312662}, \href
  {http://adsabs.harvard.edu/abs/2000ApJ...534L.135M} {534, L135}

\bibitem[\protect\citeauthoryear{{Molendi} \& {Pizzolato}}{{Molendi} \&
  {Pizzolato}}{2001}]{Molendi2001}
{Molendi} S.,  {Pizzolato} F.,  2001, \mn@doi [\apj] {10.1086/322387}, \href
  {http://adsabs.harvard.edu/abs/2001ApJ...560..194M} {560, 194}

\bibitem[\protect\citeauthoryear{{Peterson} \& {Fabian}}{{Peterson} \&
  {Fabian}}{2006}]{Peterson2006}
{Peterson} J.~R.,  {Fabian} A.~C.,  2006, \mn@doi [\physrep]
  {10.1016/j.physrep.2005.12.007}, \href
  {http://adsabs.harvard.edu/abs/2006PhR...427....1P} {427, 1}

\bibitem[\protect\citeauthoryear{{Pfrommer}}{{Pfrommer}}{2013}]{Pfrommer2013}
{Pfrommer} C.,  2013, \mn@doi [\apj] {10.1088/0004-637X/779/1/10}, \href
  {http://adsabs.harvard.edu/abs/2013ApJ...779...10P} {779, 10}

\bibitem[\protect\citeauthoryear{{Pfrommer} \& {En{\ss}lin}}{{Pfrommer} \&
  {En{\ss}lin}}{2004}]{pfrommer2004}
{Pfrommer} C.,  {En{\ss}lin} T.~A.,  2004, \mn@doi [\aap]
  {10.1051/0004-6361:20031464}, \href
  {http://adsabs.harvard.edu/abs/2004A%26A...413...17P} {413, 17}

\bibitem[\protect\citeauthoryear{{Pfrommer}, {En{\ss}lin}  \&
  {Sarazin}}{{Pfrommer} et~al.}{2005}]{Pfrommer2005}
{Pfrommer} C.,  {En{\ss}lin} T.~A.,   {Sarazin} C.~L.,  2005, \mn@doi [\aap]
  {10.1051/0004-6361:20041576}, \href
  {http://adsabs.harvard.edu/abs/2005A%26A...430..799P} {430, 799}

\bibitem[\protect\citeauthoryear{{Prokhorov} \& {Churazov}}{{Prokhorov} \&
  {Churazov}}{2014}]{Prokhorov2014}
{Prokhorov} D.~A.,  {Churazov} E.~M.,  2014, \mn@doi [\aap]
  {10.1051/0004-6361/201322454}, \href
  {http://adsabs.harvard.edu/abs/2014A%26A...567A..93P} {567, A93}

\bibitem[\protect\citeauthoryear{{Prokhorov}, {Antonuccio-Delogu}  \&
  {Silk}}{{Prokhorov} et~al.}{2010}]{Prokhorov2010}
{Prokhorov} D.~A.,  {Antonuccio-Delogu} V.,   {Silk} J.,  2010, \mn@doi [\aap]
  {10.1051/0004-6361/200913920}, \href
  {http://adsabs.harvard.edu/abs/2010A%26A...520A.106P} {520, A106}

\bibitem[\protect\citeauthoryear{{Ruszkowski}, {Yang}  \&
  {Reynolds}}{{Ruszkowski} et~al.}{2017}]{Ruszkowski2017}
{Ruszkowski} M.,  {Yang} H.-Y.~K.,   {Reynolds} C.~S.,  2017, preprint, \href
  {http://adsabs.harvard.edu/abs/2017arXiv170107441R} {} (\mn@eprint {arXiv}
  {1701.07441})

\bibitem[\protect\citeauthoryear{{Sarazin}}{{Sarazin}}{1986}]{Sarazin1986}
{Sarazin} C.~L.,  1986, \mn@doi [Reviews of Modern Physics]
  {10.1103/RevModPhys.58.1}, \href
  {http://adsabs.harvard.edu/abs/1986RvMP...58....1S} {58, 1}

\bibitem[\protect\citeauthoryear{{Selig}, {Vacca}, {Oppermann}  \&
  {En{\ss}lin}}{{Selig} et~al.}{2015}]{Selig}
{Selig} M.,  {Vacca} V.,  {Oppermann} N.,   {En{\ss}lin} T.~A.,  2015, \mn@doi
  [\aap] {10.1051/0004-6361/201425172}, \href
  {http://adsabs.harvard.edu/abs/2015A%26A...581A.126S} {581, A126}

\bibitem[\protect\citeauthoryear{{Tamura} et~al.,}{{Tamura}
  et~al.}{2001}]{Tamura2001}
{Tamura} T.,  et~al., 2001, \mn@doi [\aap] {10.1051/0004-6361:20000038}, \href
  {http://adsabs.harvard.edu/abs/2001A%26A...365L..87T} {365, L87}

\bibitem[\protect\citeauthoryear{{Vikhlinin}, {Kravtsov}, {Forman}, {Jones},
  {Markevitch}, {Murray}  \& {Van Speybroeck}}{{Vikhlinin}
  et~al.}{2006}]{Vikhlinin2006}
{Vikhlinin} A.,  {Kravtsov} A.,  {Forman} W.,  {Jones} C.,  {Markevitch} M.,
  {Murray} S.~S.,   {Van Speybroeck} L.,  2006, \mn@doi [\apj]
  {10.1086/500288}, \href {http://adsabs.harvard.edu/abs/2006ApJ...640..691V}
  {640, 691}

\bibitem[\protect\citeauthoryear{{V{\"o}lk}, {Aharonian}  \&
  {Breitschwerdt}}{{V{\"o}lk} et~al.}{1996}]{Volk1996}
{V{\"o}lk} H.~J.,  {Aharonian} F.~A.,   {Breitschwerdt} D.,  1996, \mn@doi
  [\ssr] {10.1007/BF00195040}, \href
  {http://adsabs.harvard.edu/abs/1996SSRv...75..279V} {75, 279}

\bibitem[\protect\citeauthoryear{{Wentzel}}{{Wentzel}}{1974}]{Wentzel1974}
{Wentzel} D.~G.,  1974, \mn@doi [\araa] {10.1146/annurev.aa.12.090174.000443},
  \href {http://adsabs.harvard.edu/abs/1974ARA%26A..12...71W} {12, 71}

\bibitem[\protect\citeauthoryear{{Wiener}, {Oh}  \& {Guo}}{{Wiener}
  et~al.}{2013}]{Wiener2013}
{Wiener} J.,  {Oh} S.~P.,   {Guo} F.,  2013, \mn@doi [\mnras]
  {10.1093/mnras/stt1163}, \href
  {http://adsabs.harvard.edu/abs/2013MNRAS.434.2209W} {434, 2209}

\bibitem[\protect\citeauthoryear{{Zhuravleva} et~al.,}{{Zhuravleva}
  et~al.}{2013}]{Zhuravleva2013}
{Zhuravleva} I.,  et~al., 2013, \mn@doi [\mnras] {10.1093/mnras/stt1506}, \href
  {http://adsabs.harvard.edu/abs/2013MNRAS.435.3111Z} {435, 3111}

\makeatother
\end{thebibliography}

\end{document}